\documentclass{elsart}

\usepackage{amssymb}

\begin{document}

\begin{frontmatter}



\title{Nonextensive triangle equality and other properties of Tsallis relative-entropy minimization}


\author{Ambedkar Dukkipati,\thanksref{email1}}
\author{M. Narasimha Murty,\corauthref{cor}\thanksref{email2}}
\author{Shalabh Bhatnagar\thanksref{email3}}
\corauth[cor]{corresponding author}
\thanks[email1]{ambedkar@csa.iisc.ernet.in}
\thanks[email2]{mnm@csa.iisc.ernet.in (Tel:+91-80-22932779)}
\thanks[email3]{shalabh@csa.iisc.ernet.in}
\address{Department of Computer Science and Automation,
Indian Institute of Science,\\
Bangalore-560012, India.}

\begin{abstract}
 	Kullback-Leibler relative-entropy has unique properties in 
 	cases involving distributions resulting from relative-entropy
 	minimization. Tsallis relative-entropy is a one parameter
 	generalization of Kullback-Leibler relative-entropy in the
 	nonextensive thermostatistics. In this paper, we present the properties
 	of Tsallis relative-entropy  minimization and present some
 	differences with the classical case. In the representation of
 	such a minimum relative-entropy distribution, we highlight the
 	use of the {\em $q$-product}, an operator that has
	been recently introduced to derive the mathematical structure
	behind the Tsallis statistics. One of our main results is
	generalization of {\em triangle equality} of relative-entropy
	minimization to the nonextensive case.
\end{abstract}

\end{frontmatter}

\section{Introduction}
\label{Section:Introduction}
	Maximum and minimum entropy methods, known as ME
	methods,
	originally coming from 
	physics, have been promoted to a general principle of inference
	primarily by the works of
	Jaynes~\cite{Jaynes:1983:PapersOnProbabilityStatisticsAndStatisticalPhysics}
	and later by
	Kullback~\cite{Kullback:1959:InformationTheoryAndStatistics}.
	Jaynes maximum entropy principle involves maximizing Shannon
	entropy~\cite{Elsasser:1937:OnQuantumMeasuremetnsAndTheRoleOfUncertainity,Jaynes:1957:InformationTheoryStatisticalMechanics} 
	while Kullback minimum entropy principle involves minimizing
	Kullback-Leibler
	relative-entropy~\cite{Kullback:1959:InformationTheoryAndStatistics}.
	Logarithmic form of 
	information measure is common for all these entropies. 

	On the other hand, however, Tsallis
	in~\cite{Tsallis:1988:GeneralizationOfBoltzmannGibbsStatistics}
	proposed a non-logarithmic form of entropy (termed as
	nonextensive entropy or Tsallis entropy) which is
	considered as a useful measure in describing thermostatistical
	properties of a certain class of physical 
	systems that entail long-range interactions, long-term
	memories and multi-fractal structures. The thermostatistical
	formalism based on Tsallis entropy is termed as
	generalized or nonextensive thermostatistics, since Tsallis
	entropy is a one-parameter generalization 
	of Shannon entropy and does not satisfy {\em additive}
	property involving independent probability distributions but
	satisfies the so-called {\em pseudo additivity} or {\em
	nonextensive additivity} (see
	(\ref{Equation:NonextensiveAdditivityOfTsallisEntropy})).
	These generalized statistics have been 
	applied not only to physical systems but also to various
	problems in optimization (generalized 
	simulated annealing~\cite{TsallisStariolo:1996:GSA}),
	statistical
	inference~\cite{Tsallis:1998:GeneralizedEntropyBasedCriterionForConsistentLearning},
	machine
	learning~\cite{AnastasiadisMagoulas:2004:NonextensiveStatisticalMechanicsForNN}
	etc. Recently Shannon-Khinchin  
	axioms have been 
	generalized to nonextensive
	systems~\cite{Suyari:2004:GeneralizationOfShannonKhinchinAxioms}.

	On similar lines, generalization of relative-entropy, called
	Tsallis relative-entropy, has been 
	proposed
	in~\cite{Tsallis:1998:GeneralizedEntropyBasedCriterionForConsistentLearning}. 
	In this paper, we study Tsallis relative-entropy
	minimization and its
	differences with the classical case. We generalize the
	triangle equality
	(see~(\ref{Equation:Classical_TriangleEquality})) of
	Kullback-Leibler relative-entropy 
	minimization,
	which qualifies  
	relative-entropy minimization as an optimal inference
	procedure with respect to relative-entropy as an information
	measure, to Tsallis relative-entropy case. 

	We present the necessary background in
	\S~\ref{Section:Background}, where we discuss properties of
	relative-entropy minimization in the classical case, and we give the
	basic definitions related to Tsallis entropy. In
	\S~\ref{Section:TsallisRelativeEntropyMinimization} we
	present the relative-entropy minimization in non-extensive
	framework and discuss its differences with the classical
	case. Finally, triangle equality for Tsallis relative-entropy
	minimization is derived
	in \S~\ref{Section:NonextensiveTriangleEquality}. A brief
	discussion of Tsallis relative-entropy minimization in the
	case of ``normalized $q$-expected'' values is presented in
	\S~\ref{Section:InTheCaseOfNormalized-q-expectations}. 

\section{Background}
\label{Section:Background}
  \subsection{Relative-entropy minimization: In classical case}
	Minimizing the Kullback-Leibler relative-entropy (or
	I-divergence or cross-entropy) with respect to a set of moment
	constraints finds its importance in the celebrated {\it
	Kullback's minimum relative-entropy 
	principle}~\cite{Kullback:1959:InformationTheoryAndStatistics}.
        This principle is a general method of inference about an
        unknown probability distribution when there exists a prior
        estimate of the distribution and new information in the form of
        constraints on expected
        values~\cite{Shore:1981:PropertiesOfCrossEntropyMinimization}.
        Formally, we can state the minimum relative-entropy principle as:
	given a prior distribution $r$, of all the 
        probability distributions that satisfy the given moment
	constraints, one should choose the
        \textit{posterior} $p$ with the least relative-entropy
	\begin{equation}
	\label{Equation:KullbackLeiblerEntropy}
	I_{1}(p\|r) =  \int dx \,p(x) \ln \frac{p(x)}{r(x)} \enspace,
	\end{equation}
	provided that the integral above exists\footnote{In measure theoretic
	terms, the integral exists if the measure induced by $p$ is
	{\em absolutely continuous} with respect to that 
	induced by $r$, otherwise $I_{1}(p\|r) =
	\infty$~\cite{Csiszar:1975:I-devergenceOfProbabilityDistributionsAndMinimizationProblems}. In this work we do not aim at mathematical rigor of the measure theoretic information theory. In particular, we assume that all quantities of interest exist for all distributions considered. Note that the measure theoretic definitions of these quantities relies strongly on the {\em Lebesgue-Radon-Nikodym} Theorem~\cite{Cherny:2004:OnMinimizationAndMaximizationOfEntropy}.}.     
	The prior distribution $r$ can be a 
	{\it reference} distribution (uniform, Gaussian, Lorentzian
	or Boltzmann etc.) or a {\it prior} estimate of $p$.

	The principle of Jaynes maximum
	entropy
	is a special case of minimization of relative-entropy under
	appropriate
	conditions~\cite{ShoreJohnson:1980:AxiomaticDerivationOfThePrincipleMaxEntMinEnt}.
	In particular, minimizing relative-entropy is equivalent to
	maximizing Shannon 
	entropy when the prior is a uniform distribution.
	Relative-entropy minimization has been applied
	primarily to
	statistics~\cite{Kullback:1959:InformationTheoryAndStatistics},
	and also to statistical
	mechanics~\cite{Hobson:1971:ConceptsInStatisticalMechanics_SecondaryRef},
	pattern
	recognition~\cite{ShoreCray:1982:MinimumCrossEntropyPatternClassification},
	spectral 
	analysis~\cite{Shore:1981:MinimumCrossEntropySpectralAnalysis_SecondaryRef},
	speech
	coding~\cite{MarkelGray:1976:LinearPredictionOfSpeach_SecondaryRef}.
	For a list of references on applications of relative-entropy
	minimization
	see~\cite{ShoreJohnson:1980:AxiomaticDerivationOfThePrincipleMaxEntMinEnt,Cherny:2004:OnMinimizationAndMaximizationOfEntropy}. 

	Properties of relative-entropy minimization have been studied and
	presented extensively
	in~\cite{Shore:1981:PropertiesOfCrossEntropyMinimization}.  
	Here we briefly mention a few. Given a prior distribution $r$
	with a finite set of moment constraints of the form
	\begin{equation}
	\label{Equation:Constraint2} 
	\int dx \, u_{m}(x) p(x)  = {\langle{u}_{m}\rangle}\enspace, \:\:\: m =1,
	\ldots M \enspace, 
	\end{equation}
	along with the normalizing constraint $\int p(x) dx = 1$ (from
	now on we assume that any set of constraints on probability
	distributions implicitly includes this constraint), 
	the minimum relative-entropy distribution is of the form
	\begin{equation}
	\label{Equation:MinimumCrossEntropyDistribution}  
	  p(x) = \frac{r(x) e^{-\sum_{m=1}^{M} \beta_{m} u_{m}(x)} }
	{\widehat{Z_{1}}} \enspace, 
	\end{equation}
	where
	\begin{equation}
	\widehat{Z_{1}} = \int dx \, r(x) e^{-\sum_{m=1}^{M}
	\beta_{m} u_{m}(x)  } 
	\end{equation}
	is the
	partition function, 
	$\beta_{m},\: m =1, \ldots, M$ are the corresponding Lagrange
	multipliers, $u_{m},\: m = 1, \ldots M$ are some functions of
	the underlying random variable whose expectation values
	$\langle u_{m} \rangle, \: m = 1, \ldots M $ are (assumedly)
	{\it a priori} known. When 
	the prior is a uniform distribution with a compact support of
	$W$ ($W$ possible configurations in discrete case), the minimum
	relative-entropy distribution turns out to be 
	\begin{equation}
	\label{Equation:MinimumCrossEntropyDistribution_WithUniformPrior}
	p(x) = \frac{\displaystyle e^{ -\sum_{m=1}^{M} \beta_{m}
	u_{m}(x)}}{\displaystyle \int dx \, e^{
	-\sum_{m=1}^{M} \beta_{m} u_{m}(x)} } \enspace, 
	\end{equation}
	which is in fact a maximum entropy distribution (Boltzmann
	distribution)  of Shannon entropy with respect to the
	constraints (\ref{Equation:Constraint2}).

	Many properties of relative-entropy minimization just reflect
	well-known properties of
	relative-entropy
	but there are surprising differences as
	well~\cite{Shore:1981:PropertiesOfCrossEntropyMinimization}.
	For example, 
	relative-entropy does not generally satisfy a triangle relation
	involving three arbitrary probability distributions. But in
	certain important cases involving distributions that result
	from relative-entropy minimization, relative-entropy does satisfy
	{\em triangle equality}.

	The statement of triangle equality can be formulated as follows.
	Let $r$ be the prior distribution, $p$ be the probability
	distribution that minimizes the relative-entropy subject to
	set of constraints (\ref{Equation:Constraint2}) and $l$ be any other distribution
	satisfying the same constraints, then we have the triangle
	equality~\cite{Shore:1981:PropertiesOfCrossEntropyMinimization}:
	\begin{equation}
	\label{Equation:Classical_TriangleEquality}
	I_{1}(l\|r) = I_{1}(l\|p) + I_{1}(p\|r) \enspace.
	\end{equation}
	This triangle equality is important for application in which
	relative-entropy minimization is used for purposes of pattern
	classification and cluster
	analysis~\cite{ShoreCray:1982:MinimumCrossEntropyPatternClassification}.

  \subsection{Nonextensive framework}

	Tsallis entropy, which was introduced by
	Tsallis~\cite{Tsallis:1988:GeneralizationOfBoltzmannGibbsStatistics}
	is given by
	\begin{equation}
	\label{Equation:TsallisEntropy}
	S_{q}(p) = - \int dx \, p(x) \frac{{p(x)}^{q-1}-1}{q-1}  \enspace,
	\end{equation}
	where $q \in \mathbb{R}$ is called non-extensive index ($q$ is
	positive in order to ensure the concavity of
	$S_{q}$).
	Tsallis entropy is a one-parameter generalization of Shannon
	entropy in the sense that
	\begin{equation}
	\lim_{q \to 1} S_{q}(p) = - \int dx \, p(x) \ln p(x)  = S_{1}(p) \enspace.
	\end{equation}
	The entropic index $q$ characterizes the degree of
	nonextensivity reflected in the pseudo-additivity property
	\begin{equation}
	\label{Equation:NonextensiveAdditivityOfTsallisEntropy}
	S_{q}(A+B) = S_{q}(A) + S_{q}(B) + (1-q) S_{q}(A)S_{q}(B)\enspace,
	\end{equation}
	where $A$ and $B$ are two {\em independent systems} in the
	sense that the probability distribution of $A+B$ factorizes
	into those of $A$ and $B$.

	Maximizing the Tsallis entropy $S_{q}$ with respect to the constraints
	\begin{equation}
	\label{Equation:GeneralizedConstraint2} 
	\int dx \, u_{m}(x) {p(x)}^{q}  = {\langle{u_{m}}\rangle}_{q} \enspace, \:\:\: m =1, \ldots M
	\enspace, 
	\end{equation}
	the generalized equilibrium
	probability distribution is found to
	be~\cite{TsallisMendesPlastino:1998:TheRoleOfConstraints},
	\begin{equation}
	\label{Equation:TsallisMaximumEntropyDistribution}
	 p(x) = \frac{\displaystyle \left[1 - (1-q) \sum_{m=1}^{M} \beta_{m}
	u_{m}(x) \right]^{\frac{1}{1-q}}}{\displaystyle Z_{q}}  \enspace, 
	\end{equation}
	where
	\begin{equation}
	Z_{q} = \int dx \left[1-(1-q) \sum_{m=1}^{M} \beta_{m}
	u_{m}(x) \right]^{\frac{1}{1-q}} 
	\end{equation}
	is the partition function,
	and $\beta_{m}, \:m = 1, \ldots M$ are the corresponding Lagrange
	multipliers\footnote{To avoid proliferation of symbols we use
	same notation for the minimum or 
	maximum entropy distributions and Lagrange multipliers in the 
	various cases; the correspondence should be clear from the
	context.} and ${\langle u_{m} \rangle}_{q}$ is a known
	{\em $q$-expectation} of
	$u_{m}$~\cite{TsallisMendesPlastino:1998:TheRoleOfConstraints}. 
	This distribution is called generalized maximum entropy distribution
	or simply Tsallis 
	distribution~\cite{PlastinoPlastino:1994:FromGibbsToTsallisCanonicalDistribution}. 
	The limit $q \rightarrow 1$ in
	(\ref{Equation:TsallisMaximumEntropyDistribution}), recovers
	the maximum entropy distribution in the classical case. 

	Now, the definition of Kullback-Leibler relative-entropy
	$I_{1}$~(\ref{Equation:KullbackLeiblerEntropy}) and the
	generalized entropic form
	$S_{q}$~(\ref{Equation:TsallisEntropy}) naturally lead to the 
	generalization~\cite{Tsallis:1998:GeneralizedEntropyBasedCriterionForConsistentLearning}
	\begin{equation}
	\label{Equation:GeneralizedKullbackLeiblerEntropy}
	I_{q}(p\|r) = \int dx \, p(x) \frac{\displaystyle
	{\left[\frac{p(x)}{r(x)}\right]}^{q-1} -1 }{\displaystyle q-1}
	 \enspace,
	\end{equation}
	which is called as
	Tsallis relative-entropy.
	The limit $q \rightarrow 1$ recovers the 
	relative-entropy in the classical case. Also one can verify 
	that (see~\cite{Tsallis:1998:GeneralizedEntropyBasedCriterionForConsistentLearning} ) 
	\begin{eqnarray}
	I_{q}(p\|r) & \geq & 0 \:\:\: \textrm{if $ q > 0$} \nonumber\\
	            &   =  & 0 \:\:\: \textrm{if $ q = 0$} \nonumber\\
	            & \leq & 0 \:\:\: \textrm{if $ q < 0$} \enspace.
	\end{eqnarray}
	For $q\neq 0$, the equalities hold if and only if $p=r$ almost
	everywhere. Further, for $q >0$, $I_{q}(p\|r)$ is a convex
	function of $p$ and $r$, and for $q<0$ it is
	concave~\cite{BorlandPlastinoTsallis:1998:InformationGainWithinNonextensiveThermostatistics}.
	Like Tsallis entropy, Tsallis relative-entropy satisfies the
	pseudo-additivity of the form~\cite{FuruichiYanagiKuriyama:2004:FundamentalPropertiesOfTsallisRelativeEntropy}
	{\setlength\arraycolsep{0pt}
	\begin{eqnarray}
	\label{Equation:NonextensiveAdditivityOfTsallisRelativeEntropy}
	I_{q}(A_{1} + A_{2} \| B_{1} + B_{2}) = I_{q}(A_{1} \| B_{1})
	&&+ I_{q}( A_{2} \| B_{2}) \nonumber \\ && + (q-1) I_{q}(A_{1}
	\| B_{1}) I_{q}( A_{2} \| B_{2}) \enspace, 
	\end{eqnarray}}
	where $A_{1}$, $A_{2}$ and $B_{1}$, $B_{2}$ are the
	independent pairs.
	The limit $q \rightarrow 1$ in
	(\ref{Equation:NonextensiveAdditivityOfTsallisRelativeEntropy})
	retrieves
	\begin{equation}
	\label{Equation:ClassicalAdditivityOfRelativeEntropy}
	I_{1}(A_{1} + A_{2} \| B_{1} + B_{2}) = I_{1}(A_{1} \| B_{1})
	+ I_{1}( A_{2} \| B_{2}) \enspace
	\end{equation}
	the additivity property of Kullback-Leibler relative-entropy.
	
	Further properties of Tsallis relative-entropy have
	been discussed in
	\cite{Tsallis:1998:GeneralizedEntropyBasedCriterionForConsistentLearning,BorlandPlastinoTsallis:1998:InformationGainWithinNonextensiveThermostatistics,FuruichiYanagiKuriyama:2004:FundamentalPropertiesOfTsallisRelativeEntropy}.
	Characterization of Tsallis relative-entropy, by generalizing
	Hobson's uniqueness
	theorem~\cite{Hobson:1969:AnewTheoremOfInformationTheory} of
	relative-entropy, is presented
	in~\cite{Furuichi:2004:AcharacterizationOfTheTsallisRealtiveEntropy}.

\section{Tsallis relative-entropy minimization}
\label{Section:TsallisRelativeEntropyMinimization}

 \subsection{Generalized minimum relative-entropy distribution}

	To minimize Tsallis relative-entropy with respect to
	the set of constraints
	(\ref{Equation:GeneralizedConstraint2}) the concomitant 
	variational principle can be written as
	\begin{displaymath}
 	\delta \left\{
	  I_{q}(p\|r) 
         + \lambda \left( \int dx\, p(x) -1 \right) 
	 + \sum_{m=1}^{M} \beta_{m} \left( \int dx\, {p(x)}^{q}
 	u_{m}(x)   - {\langle{{u}_{m}}\rangle}_{q} \right) \right\} = 0
 	\enspace,
	\end{displaymath}
	where $\lambda$ and $\beta_{m}, \:\: m=1, \ldots M$ are
	Lagrange multipliers.
	This gives us minimum Tsallis relative-entropy distribution
	as~\cite{BorlandPlastinoTsallis:1998:InformationGainWithinNonextensiveThermostatistics} 
	 \begin{equation}
	 \label{Equation:GeneralizedMinimumCrossEntropyDistribution_1}  
	  p(x) = \frac{\displaystyle {\left[{r(x)}^{1-q} - (1-q)
	 \sum_{m=1}^{M} \beta_{m} u_{m}(x)
	 \right]}^{\frac{1}{1-q}}}{\displaystyle 
	\widehat{Z_{q}}} \enspace,   
	 \end{equation}
	where values of $\beta_{m},\:\: m = 1 \ldots M$ are determined
 	by the constraints 
	(\ref{Equation:GeneralizedConstraint2}) and $\widehat{Z_{q}}$,
	the partition function, is given by
	\begin{equation}
	\widehat{Z_{q}} = \int dx {\left[{r(x)}^{1-q} - (1-q)
	 \sum_{m=1}^{M} \beta_{m} u_{m}(x)  \right]}^{\frac{1}{1-q}}
	 \enspace. 
	\end{equation}

	A note on the constraint
	(\ref{Equation:GeneralizedConstraint2}). This constraint had
	been used for some
	time~\cite{CuradoTsallis:1991:GSMconnectionsWithThermodynamics},
	but because of problems in justifying it  on physical grounds 
	the constraint
	\begin{equation}
	\label{Equation:EscortConstraint}
	\frac{\displaystyle \int dx\,{p(x)}^{q}u_{m}(x)
	}{\displaystyle \int dx\, {p(x)}^{q}} = {\langle\langle u_{m} \rangle\rangle}_{q}
	\end{equation}
	has been introduced
	in~\cite{TsallisMendesPlastino:1998:TheRoleOfConstraints}.
	${\langle\langle u_{m} \rangle\rangle}_{q}$ is called {\em normalized
	q-expectation value} of $u_{m}$.
	We discuss Tsallis
	relative-entropy minimization and its properties with respect
	to the constraint 
	(\ref{Equation:EscortConstraint}), briefly, in
	\S~\ref{Section:InTheCaseOfNormalized-q-expectations}.  

  \subsection{q-product representation of generalized minimum
  relative-entropy distribution}

	The mathematical basis for Tsallis statistics comes from the
	q-deformed expressions for the logarithm
	(\textit{$q$-logarithm}) and the exponential function
	(\textit{$q$-exponential}) which were first proposed
	in~\cite{Tsallis:1994:WhatAreTheNumbersThatExperimentsProvide},
	in the context of nonextensive thermostatistics.
	The $q$-logarithm is defined as
	\begin{equation}
	\ln_{q} x = \frac{\displaystyle x^{1-q} -1}{\displaystyle 1-q}
	\:\:\: (x >0, q \in \mathbb{R}),
	\end{equation}
	and the $q$-exponential is defined as
	 \begin{equation}
	 \label{Equation:qExponential}
	 e_{q}^{x} = \left\{ \begin{array}{ll}
	   {[1+(1-q)x]}^{\frac{1}{1-q}} & \textrm{if $ 1 + (1-q)x \geq 0 $} \\
	   0 & \textrm{otherwise.}  
	   \end{array} \right. 
	 \end{equation}
	These two functions are related by
	\begin{equation}
	\label{Equation:Relation_q-exponentialandLogorithm}
	e_{q}^{\ln_{q} x} = x \enspace.
	\end{equation}
	Properties of these q-deformed functions are studied 
	in~\cite{Yamano:2002:SomePropertiesOfq-logrithmAndq-exponential}. 
	In this framework a new multiplication operation, called $q$-product
	\begin{equation}
	\label{Equation:q-product}
	x \otimes_{q} y \equiv \left\{ \begin{array}{ll}
	\left( x^{1-q} + y^{1-q} -1 \right)^{\frac{1}{1-q}} &
	\textrm{if $x,y >0$ and}\:\: 
	     \textrm{$x^{1-q} + y^{1-q} -1 > 0$}\\
          0 & \textrm{otherwise.}
	   \end{array} \right.
	\end{equation}
	is first introduced 
	 in~\cite{NivanenMehauteWang:2003:GeneralizedAlgebraWithinNonextensiveStatistics}
	 and explicitly defined
	 in~\cite{Borges:2004:ApossibleDeformedAlgebra} for satisfying
	the following equations:
	\begin{eqnarray}
	\label{Equation_2}
	 \ln_{q} (x \otimes_{q} y) & = & \ln_{q} x + \ln_{q} y
	\label{Equation:q-logorithmProduct} \enspace,\\
	 e_{q}^{x} \otimes_{q} e_{q}^{y} & = & e_{q}^{x+y} \label{Equation:q-exponentialProduct} \enspace.
	\end{eqnarray}
	The $q$-product recovers the usual product in the limit $q
	\rightarrow 1$ i.e., 
	 $\lim_{q \to 1} (x \otimes_{q} y) = xy $. The fundamental
	 properties of the $q$-product $\otimes_{q}$ are almost the
	 same as the usual product, and in general 
	\begin{displaymath}
	a (x \otimes_{q} y ) \neq ax \otimes_{q} y \:\:\: (a,x,y \in \mathbb{R})\enspace.
	\end{displaymath}
	Further properties of the $q$-product can be found in
	 \cite{NivanenMehauteWang:2003:GeneralizedAlgebraWithinNonextensiveStatistics,Borges:2004:ApossibleDeformedAlgebra}.
	 Also, $q$-product has been used in various applications of
	Tsallis
	statistics~\cite{SuyariTsukada:2005:LawOfErrorInTsallisStatistics}. 

	Previously, generalized entropies and maximum entropy distributions have
	been represented in terms of $q$-logarithm and
	$q$-exponential. We list some of them which we are going to
	use later in this paper.
	Tsallis entropy (\ref{Equation:TsallisEntropy}) can be
	represented as
	\begin{equation}
	\label{Equation:TsallisEntropy_q-LogorithmForm}
	S_{q}(p) = - \int dx\, {p(x)}^{q} \ln_{q} p(x)   \enspace,
	\end{equation}
	and Tsallis relative-entropy
	(\ref{Equation:GeneralizedKullbackLeiblerEntropy}) as 
	\begin{equation}
	\label{Equation:GeneralizedKullbackLeibler_q-LogorithmForm}
	I_{q}(p\|r) = - \int dx\, p(x) \ln_{q}\frac{r(x)}{p(x)} \enspace.
	\end{equation}
	One can represent Tsallis
	distribution ~(\ref{Equation:TsallisMaximumEntropyDistribution})
	in terms of {\em $q$-exponential} as 
	\begin{equation}
	\label{Equation:TsallisDistributionInq-exponentialForm}
	p(x) = \frac{e_{q}^{ - \sum_{m=1}^{M} \beta_{m} u_{m}(x)}}{Z_{q}} \enspace,
	\end{equation}
	since $p(x)=0$ whenever
	$ \left[ 1 - (1-q) \sum_{m=1}^{M} \beta_{m} u_{m}(x) \right] <
	0 $ which is {\em Tsallis cut-off
	condition}~\cite{Tsallis:1988:GeneralizationOfBoltzmannGibbsStatistics}
	assumed implicitly. 

	Note that generalized relative-entropy distribution
	 (\ref{Equation:GeneralizedMinimumCrossEntropyDistribution_1}),
	 is not of the form  
	 of~(\ref{Equation:MinimumCrossEntropyDistribution}) even if we
	 replace the exponential with the q-exponential. 
	 But one can verify the non-trivial fact that 
	(\ref{Equation:GeneralizedMinimumCrossEntropyDistribution_1})
	can be expressed in a similar form as in the classical case by
	 invoking q-product as, 
	 \begin{equation}
	 \label{Equation:GeneralizedMinimumCrossEntropyDistribution_2}  
	 p(x) = \frac{\displaystyle r(x) \otimes_{q} e_{q}^{-
	 \sum_{m=1}^{M} \beta_{m} u_{m}(x)  }}{\displaystyle
	 \widehat{Z_{q}}}  \enspace, 
	 \end{equation}
	where
	\begin{displaymath}
	\widehat{Z_{q}} = \int dx\, r(x) \otimes_{q}
	 e_{q}^{ - \sum_{m=1}^{M} \beta_{m} u_{m}(x) } .
	\end{displaymath}
	One can see from the later parts of this paper, this representation is
	useful in deriving properties of Tsallis relative-entropy
	minimization.

	It is important to note that the distribution in
 	(\ref{Equation:GeneralizedMinimumCrossEntropyDistribution_1})
 	could be a (local/global) 
 	minimum only if $q > 0$ and Tsallis cut-off condition is
 	extended to the relative-entropy case i.e., $p(x) = 0$ whenever 
	$\left[{r(x)}^{1-q} - (1-q)
 	\sum_{m=1}^{M} \beta_{m} u_{m}(x) \right] < 0$.

  \subsection{Properties}

	As we mentioned earlier, in the classical case i.e., when $q=1$, 
	 relative-entropy minimization with uniform distribution as a prior
	 is equivalent to entropy maximization.
	But, in the case of nonextensive framework, 
	 this is not true. Let $r$ be the uniform distribution with
	 compact support $W$ over $E \subset \mathbb{R}$.
	Then, by
	 (\ref{Equation:GeneralizedMinimumCrossEntropyDistribution_1})
	 one can verify that probability distribution which 
	 minimizes Tsallis relative-entropy is
	\begin{displaymath}
	  p(x)  = \frac{ \displaystyle {\left[ \frac{1}{{W}^{1-q}} - (1-q)   \sum_{m=1}^{M} \beta_{m}
	u_{m}(x) \right]}^{\frac{1}{1-q}}}{ \displaystyle \int_{E} dx\, 
	{\left[\frac{1}{{W}^{1-q}}  - 
	 (1-q) \sum_{m=1}^{M} \beta_{m} 
	u_{m}(x) \right]}^{\frac{1}{1-q}}} \enspace,
	\end{displaymath}
	which can be written as (by
	 (\ref{Equation:GeneralizedMinimumCrossEntropyDistribution_3})
	 and (\ref{Equation:PropertyOflnq(x/y)})) 
	\begin{equation}
	\label{Equation:Tsallis-MinimumCrossEntropyDistribution-UniformPrior-Reviewer}	  
	p(x)  = \frac{ \displaystyle e_{q}^{ -W^{q-1} \ln_{q} W - \sum_{m=1}^{M} \beta_{m}
	u_{m}(x)}}{\displaystyle \int_{E} dx\, e_{q}^{ -W^{q-1} \ln_{q} W  - \sum_{m=1}^{M} \beta_{m}
	u_{m}(x)}  } 
	\end{equation}
	or (by the definition of $q$-exponential (\ref{Equation:qExponential}))
	\begin{equation}
	\label{Equation:Tsallis-MinimumCrossEntropyDistribution-UniformPrior}
	  p(x)  = \frac{ \displaystyle e_{q}^{- 
	W^{1-q} \sum_{m=1}^{M} \beta_{m}
	u_{m}(x)}}{\displaystyle \int_{E} dx\, e_{q}^{- W^{1-q} \sum_{m=1}^{M} \beta_{m}
	u_{m}(x)} } \enspace.
	\end{equation}
	By comparing
	(\ref{Equation:Tsallis-MinimumCrossEntropyDistribution-UniformPrior-Reviewer})
	or (\ref{Equation:Tsallis-MinimumCrossEntropyDistribution-UniformPrior})
	with Tsallis maximum entropy distribution (\ref{Equation:TsallisDistributionInq-exponentialForm})
	one can conclude (formally one can verify this by
	thermodynamical equations of Tsallis
	entropy~\cite{Tsallis:1988:GeneralizationOfBoltzmannGibbsStatistics})
	that minimizing relative-entropy is {\em not 
	equivalent}\footnote{For fixed $q$-expected values ${\langle
	u_{m} \rangle}_{q}$, the two distributions,
	 (\ref{Equation:Tsallis-MinimumCrossEntropyDistribution-UniformPrior})
	 and (\ref{Equation:TsallisDistributionInq-exponentialForm})
	 are equal, but the 
	values of corresponding Lagrange multipliers are different when
	$q\neq 1$ (while in the
	classical case they remain same). Further,
	(\ref{Equation:Tsallis-MinimumCrossEntropyDistribution-UniformPrior})
	offers the relation between the Lagrange parameters in these
	two cases. Let $\beta_{m}^{(S)},\: m = 1, \ldots M$ corresponds
	to the Lagrange parameters corresponds to the generalized maximum
	entropy distribution while $\beta_{m}^{(I)},\: m = 1, \ldots M$
	corresponds to generalized minimum relative-entropy distribution
	with uniform prior. Then, we have relation $\beta_{m}^{(S)} =
	W^{1-q}\beta_{m}^{(I)},\: m = 1, \ldots M$. } to maximizing entropy when 
	the prior is uniform distribution. The key observation here is
	 $W$ appeares in
	 ~(\ref{Equation:Tsallis-MinimumCrossEntropyDistribution-UniformPrior})
	 unlike~(\ref{Equation:TsallisDistributionInq-exponentialForm}).

	Also minimum Tsallis relative-entropy satisfies~\cite{BorlandPlastinoTsallis:1998:InformationGainWithinNonextensiveThermostatistics}:
	\begin{equation}
	\label{Equation:GeneralizedKL_ThermodyamicEquation_2}
	I_{q}(p\|r) = - \ln_{q} \widehat{Z_{q}} - \sum_{m=1}^{M}
	\beta_{m} {\langle{{u}_{m}}\rangle}_{q}  \enspace
	\end{equation}
	to prove which, one can escape lengthy algebraic
	manipulations by using $q$-deformed representations of various
	formulae.

	The thermodynamic equations for the minimum Tsallis relative-entropy are
	\begin{eqnarray}
	\frac{\partial}{\partial \beta_{m}} \ln_{q} \widehat{Z_{q}} & = &-
	{\langle{{u}_{m}}\rangle}_{q} \enspace,\:\:\: m = 1, \ldots M
	\enspace,\label{Equation:GeneralizedKL_ThermodyamicEquation_1}
	\\
	\frac{\partial I_{q}}{\partial {\langle{{u}_{m}}\rangle}_{q}
	} & = & - \beta_{m} \enspace, 
	\:\:\: m =1, \ldots M \enspace,
	\label{Equation:GeneralizedKL_ThermodyamicEquation_3} 
	\end{eqnarray}
	which generalize thermodynamic equations in the classical case. One
	should note that these thermodynamic 
	equations were proved to hold true for, essentially, any
	entropic
	measure~\cite{PlastinoPlastino:1997:OnTheUniversalityOfThermodynamics,Mendes:1997:SomeGeneralRelationsInArbitraryThermostatistics}.  

\section{Nonextensive triangle equality}
\label{Section:NonextensiveTriangleEquality}
	Before we derive equivalent of triangle equality in the
	nonextensive thermostatistics we shall discuss the
	significance of triangle equality of Kullback-Leibler
	relative-entropy minimization. 
	Significance of triangle equality comes in the following
	scenario. Let $r$ be the prior estimate of the unknown
	probability distribution $l$ about which information in the
	form of expected value constraints
	\begin{equation}
	\int dx \, u_{m}(x) l(x)   =  {\langle{u}_{m}\rangle}\enspace, \:\:\: m =1, \ldots M
	\end{equation}
	is available for fixed functions $u_{m},\: m = 1, \ldots
	M$. The problem 
	is to choose a posterior estimate $p$
	that is in some sense the best estimate of $l$ given the
	available information i.e., prior and the information in the
	form of expected values. The principle of minimum entropy
	provides a general  
	solution to this inference problem and provides us the
	estimate~(\ref{Equation:MinimumCrossEntropyDistribution}).
	Further, from triangle
	equality~(\ref{Equation:Classical_TriangleEquality}), the
	minimum relative-entropy posterior estimate of 
	$l$ is not only logically consistent, but also closer to $l$, in
	the relative-entropy sense, than is the prior $r$. Moreover,
	the difference $I_{1}(l\|r) - I_{1}(l\|p)$ is exactly the
	relative-entropy $I_{1}(p\|r)$ between the posterior and the
	prior. Hence $I_{1}(p\|r)$ can be interpreted as the amount of
	information provided by the constraints that is not inherent
	in $r$.

	Additional justification to use minimum
	relative-entropy estimate of $p$ with respect to 
	constraints (\ref{Equation:Constraint2}) is provided by the
	following {\em expected value
	matching} property~\cite{Shore:1981:PropertiesOfCrossEntropyMinimization}.
	For fixed functions $u_{m}, \: m = 1, \ldots M$, let the
	actual unknown distribution $l$ satisfy
	\begin{equation}
	\label{Equation:Constraint2_ForActualDistribution_Classical} 
	\int dx\, u_{m}(x) l(x)  =  {\langle{w}_{m}\rangle} \enspace,\:\:\: m =1, \ldots M
	\enspace.
	\end{equation}
	Now, as ${\langle u_{m} \rangle}, \:m=1,
	\ldots M$ vary, $I_{q}(l\|p)$ has the minimum value when
	\begin{equation}
	\label{Equation:ExpectedValueMatching}
	{\langle u_{m} \rangle} = {\langle w_{m} \rangle} \enspace,\:\:\: m = 1,
	\dots M.
	\end{equation}
	For the proof of expected value matching property
	see~\cite{Shore:1981:PropertiesOfCrossEntropyMinimization}.
	This property states that for a distribution $p$ of the
	form~(\ref{Equation:MinimumCrossEntropyDistribution}),
	$I_{1}(l\|p)$ is the smallest when the expected values of $p$ match
	those of $l$. In particular, $p$ is not only the
	distribution that minimizes $I_{1}(p\|r)$ but also is the
	distribution of the
	form~(\ref{Equation:MinimumCrossEntropyDistribution}) that
	minimizes $I_{1}(l\|p)$. This property is a generalization of a
	property of orthogonal
	polynomials~\cite{Geronimus:1961:OrthogonalPolynomials_SecondaryRef}
	which in the case of speech
	analysis~\cite{MarkelGray:1976:LinearPredictionOfSpeach_SecondaryRef}
	is called the ``correlation matching property''.

	From the above discussion, it is clear that to derive a triangle
	equality of Tsallis relative-entropy minimization, one should
	first deduce the equivalent of expectation matching property
	in the nonextensive case.
	Let $l$ be the actual unknown distribution which satisfies 
	\begin{equation}
	\label{Equation:Constraint2_ForActualDistribution} 
	\int dx\, u_{m}(x) {l(x)}^{q}   =  {\langle{w}_{m}\rangle}_{q} \enspace,
	\:\:\: m =1, \ldots M 
	\enspace,
	\end{equation}
	$r$
	be the prior estimate of $l$ and $p$ be the posterior which
	satisfies constraints (\ref{Equation:GeneralizedConstraint2}).
	That is, we would like to find the values of
	${\langle{u}_{m}\rangle}_{q}$ for which $I_{q}(l,p)$ is
	minimum. We write the following useful relations before we proceed
	to the derivation.

	We can write generalized minimum relative-entropy distribution
	(\ref{Equation:GeneralizedMinimumCrossEntropyDistribution_2})
	as
	 \begin{equation}
	\label{Equation:GeneralizedMinimumCrossEntropyDistribution_3}   
	 p(x) = \frac{\displaystyle e_{q}^{\ln_{q} r(x)} \otimes_{q}
	{e_{q}}^{- \sum_{m=1}^{M} \beta_{m} u_{m}(x)  }}{\displaystyle
	\widehat{Z_{q}}}  
	= \frac{\displaystyle {e_{q}}^{- \sum_{m=1}^{M} \beta_{m}
	u_{m}(x) + \ln_{q} r(x) }}{\displaystyle \widehat{Z_{q}} }
	\enspace, 
	 \end{equation}
	by (\ref{Equation:Relation_q-exponentialandLogorithm}) and (\ref{Equation:q-exponentialProduct}).
	Further by using
	\begin{equation}
	\ln_{q}(xy) = \ln_{q} x + \ln_{q} y + (1-q) \ln_{q} x
	\ln_{q}y
	\end{equation}
	we get the relation
	\begin{equation}
	\label{Equation:IntermediateRelationOfGeneralizedMinimumCrossEntropyDistribution}
	\ln_{q} p(x) + \ln_{q} \widehat{Z_{q}} +  (1-q)
	\ln_{q} p(x) \ln_{q} \widehat{Z_{q}} =  
	- \sum_{m=1}^{M} \beta_{m} u_{m}(x) + \ln_{q} r(x) \enspace.
	\end{equation}
	By the property of
	$q$-logarithm~\cite{Furuichi:2004:ChainRulesAndSubadditivitiesForTsallisEntropies}  
	\begin{equation}
	\label{Equation:PropertyOflnq(x/y)}
	\ln_{q}\left(\frac{x}{y} \right) = y^{q-1}( \ln_{q}x -
	\ln_{q}y) \enspace.
	\end{equation}
	and by (\ref{Equation:TsallisEntropy_q-LogorithmForm}),
	(\ref{Equation:GeneralizedKullbackLeibler_q-LogorithmForm})
	one can verify that
	\begin{equation}
	\label{Equation:RelationBetweenGeneralizedKLAndTsallis}
	I_{q}(p\|r) = - \int dx\, {p(x)}^{q} \ln_{q}r(x)   - S_{q}(p) \enspace.
	\end{equation}

	To proceed with the derivation, consider
	\begin{displaymath}
	I_{q}(l\|p) = - \int dx\, l(x) \ln_{q} \frac{p(x)}{l(x)}  \enspace.
	\end{displaymath}
	By (\ref{Equation:PropertyOflnq(x/y)}) we have
	{\setlength\arraycolsep{1pt}
	\begin{eqnarray}
	I_{q}(l\|p) & = & - \int dx\, {l(x)}^{q} \left[ \ln_{q} p(x) -
	\ln_{q} l(x) \right]   \nonumber \\
		   & = & I_{q}(l\|r) - \int dx\, {l(x)}^{q}\left[\ln_{q}
	p(x) - \ln_{q} r(x) \right]  \enspace. \nonumber \\  
	\end{eqnarray}}
	From
	(\ref{Equation:IntermediateRelationOfGeneralizedMinimumCrossEntropyDistribution}),
	we get
	{\setlength\arraycolsep{0pt}
	\begin{eqnarray}
	I_{q}(l\|p) = I_{q}(l\|r)
	+ \int dx\, {l(x)}^{q} && \left( \sum_{m=1}^{M} \beta_{m} u_{m}(x)
	\right)   
	 + \ln_{q} \widehat{Z_{q}} \int dx\, {l(x)}^{q}  \nonumber \\
	&&+ (1-q) \ln_{q}\widehat{Z_{q}} \int dx\, {l(x)}^{q} \ln_{q} p(x)  \enspace.
	\end{eqnarray}}
	By using (\ref{Equation:Constraint2_ForActualDistribution})
	and 
	(\ref{Equation:RelationBetweenGeneralizedKLAndTsallis}), 
	{\setlength\arraycolsep{0pt}
	\begin{eqnarray}
	I_{q}(l\|p)  =  I_{q}(l\|r) + \sum_{m=1}^{M} \beta_{m}
	{\langle{{w}_{m}}\rangle}_{q} && + \ln_{q}
	\widehat{Z_{q}} \int dx\,{l(x)}^{q} \nonumber \\ 
	&& + (1-q) \ln_{q} \widehat{Z_{q}}
	\left[ - I_{q}(l\|p) - S_{q}(l) \right] \enspace,  
	\end{eqnarray}}
	and by~(\ref{Equation:TsallisEntropy}) we have
	\begin{equation}
	\label{Equation:MainIntermediateEqForNonextensiveTriangleEquality}
	I_{q}(l\|p) = I_{q}(l\|r)+ \sum_{m=1}^{M} \beta_{m}
	{\langle{{w}_{m}}\rangle}_{q} + \ln_{q} \widehat{Z_{q}}
	- (1-q) \ln_{q} \widehat{Z_{q}} I_{q}(l\|p) \enspace.
	\end{equation}
	Since the multipliers $\beta_{m},\:\ m=1,\ldots M $ are
	functions of the expected values ${\langle u_{m}
	\rangle}_{q}$, variations in the expected values are
	equivalent to variations in the multipliers. Hence to find the
	minimum of $I_{q}(l,p)$, we solve
	\begin{displaymath}
	\frac{\partial}{\partial \beta_{m}} I_{q}(l\|p) =0 \enspace,
	\end{displaymath}
	which gives us
	\begin{equation}
	\label{Equation:ClosedForm_Tsallis_ExpectedValueMatching}
	{\langle u_{m} \rangle}_{q} = \frac{ {\langle w_{m}
	\rangle}_{q} }{ 1 - (1-q) I_{q}(l\|p) } \enspace,\:\:\: m =1,
	\ldots M \enspace. 
	\end{equation}

	In the limit $q \rightarrow 1$ the above equation gives $
	{\langle u_{m} \rangle}_{1} = {\langle w_{m} \rangle}_{1}$
	which is the expectation matching property in the classical
	case. 

	Now, to derive the triangle equality for Tsallis
	relative-entropy minimization, we substitute the expression
	for ${\langle w_{m} \rangle}_{q}$,
	which is given by~(\ref{Equation:ClosedForm_Tsallis_ExpectedValueMatching}),
	in~(\ref{Equation:MainIntermediateEqForNonextensiveTriangleEquality}).
	And after some algebra one can arrive at
	\begin{equation}
	\label{Equation:NonextensiveTriangleEquality}
	I_{q}(l\|r) = I_{q}(l\|p) + I_{q}(p\|r) + (q-1) I_{q}(l\|p)
	I_{q}(p\|r) \enspace.
	\end{equation}
	The limit $q \rightarrow 1$
	in~(\ref{Equation:NonextensiveTriangleEquality}) gives the 
	triangle equality in the classical
	case~(\ref{Equation:Classical_TriangleEquality}).
	The two important cases which arise out
	of~(\ref{Equation:NonextensiveTriangleEquality}) are, 
	\begin{eqnarray}
	I_{q}(l\|r) &\leq& I_{q}(l\|p) + I_{q}(p\|r) \:\:\:
	\mbox{when} \:  0 < q \leq 1 \enspace,\\
	I_{q}(l\|r) &\geq& I_{q}(l\|p) + I_{q}(p\|r) \:\:\:
	\mbox{when} \: 1 < q \enspace.
	\end{eqnarray}
	We call~(\ref{Equation:NonextensiveTriangleEquality}) as
	{\em nonextensive triangle equality}, whose pseudo additivity
	is consistant with the pseudo additivity of Tsallis relative-entropy (compare
	(\ref{Equation:NonextensiveAdditivityOfTsallisRelativeEntropy})
	and (\ref{Equation:ClassicalAdditivityOfRelativeEntropy})),
	and hence is a natural
	generalization of triangle equality in the classical case.

\section{In the case of `normalized $q$-expectations'}
\label{Section:InTheCaseOfNormalized-q-expectations}
	In this Section we discuss Tsallis relative-entropy
	minimization with respect
	to the constraints in the form of 
	normalized $q$-expectations  
	(\ref{Equation:EscortConstraint}). For a complete discussion
	on choice of constraints
	(\ref{Equation:GeneralizedConstraint2}) and
	(\ref{Equation:EscortConstraint}) for Tsallis entropy
	maximization 
	see~\cite{TsallisMendesPlastino:1998:TheRoleOfConstraints,MartinezNicolasPenniniPlastino:2000:TsallisEntropyMaximizaionProcedureRevisited}.

	The variational principle for Tsallis
	relative-entropy minimization with respect to
	(\ref{Equation:EscortConstraint}) can be written as 
	\begin{displaymath}
 	\delta \left\{
	I_{q}(p\|r)
         + \lambda \left( \int dx\, p(x)  -1 \right) 
	 + \sum_{m=1}^{M} \beta_{m} \left(  \frac{\int dx\, {p(x)}^{q} u(x) 
	}{\int dx\, {p(x)}^{q}} -
	{\langle\langle{{u}_{m}}\rangle\rangle}_{q} \right) \right\} = 0 
 	\enspace,
	\end{displaymath}
	where $\lambda$ and $\beta_{m},\: m = 1, \ldots M $ are
	Lagrange multipliers. This gives generalized minimum relative-entropy distribution as 
	\begin{equation}
	 \label{Equation:Tsallis_minimumRelativeEntropyDistribution_For_Normalizedq-Expectations}  
	p(x) = \frac{ \displaystyle\left[{r(x)}^{1-q} - (1-q)
	\frac{\sum_{m=1}^{M} \beta_{m} \left( u_{m}(x) -
	{\langle\langle {u}_{m} \rangle\rangle}_{q} \right)}{\int dx\, {p(x)}^{q}}
	\right]^{\frac{1}{1-q}}}
	 {\widehat{\overline{{Z}_{q}}}} \enspace,  
	\end{equation}
	where
	\begin{equation}
	\widehat{\overline{{Z}_{q}}} = \int dx\, \left[{r(x)}^{1-q} - (1-q)
	\frac{\sum_{m=1}^{M} \beta_{m} \left( u_{m}(x) -
	{\langle\langle {u}_{m} \rangle\rangle}_{q} \right)}{\int dx\, {p(x)}^{q}}
	\right]^{\frac{1}{1-q}} \enspace.
	\end{equation}
	$q \rightarrow 1$
	in~(\ref{Equation:Tsallis_minimumRelativeEntropyDistribution_For_Normalizedq-Expectations})
	retrieves the minimum relative-entropy distribution in the classical case. 

	This can be expressed as
	\begin{equation}
	p(x) = \frac{r(x) \otimes_{q} e_{q}^{- \sum_{m=1}^{M}
	\beta'_{m} \left( u_{m}(x) - {\langle\langle {u}_{m}
	\rangle\rangle}_{q} \right) }}{\widehat{\overline{{Z}_{q}}}}
	\enspace, 
	\end{equation}
	where\footnote{Note that unlike Tsallis entropy
	  case~\cite{TsallisMendesPlastino:1998:TheRoleOfConstraints},
	  $\int dx\, {p(x)}^{q} \neq {\widehat{\overline{{Z}_{q}}}}^{1-q}$}
	\begin{equation}
	\beta'_{m} = \frac{\beta_{m}}{\int dx\, {p(x)}^{q}} \enspace, \:\:\: m =1,
	\ldots M \enspace.
	\end{equation}

	Minimum Tsallis relative-entropy in this case satisfies
	\begin{equation}
	I_{q}(p \| r) = - \ln_{q}\widehat{\overline{{Z}_{q}}} \enspace,
	\end{equation}
	while corresponding thermodynamical equations can be written
	as 
	\begin{eqnarray}
	\frac{\partial}{\partial \beta_{m}} \ln_{q} \widehat{Z_{q}} & = &-
	{\langle\langle{{u}_{m}}\rangle\rangle}_{q} \enspace, \:\:\: m = 1, \ldots M
	\enspace,
	\\
	\frac{\partial I_{q}}{\partial
	{\langle\langle{{u}_{m}}\rangle\rangle}_{q}  } & = & -
	\beta_{m} \enspace, \:\:\: m =1, \ldots M \enspace,
	\end{eqnarray}
	where
	\begin{equation}
	\ln_{q} \widehat{Z_{q}} = \ln_{q} \widehat{\overline{{Z}_{q}}}
	- \sum_{m=1}^{M} \beta_{m}
	{\langle\langle{{u}_{m}}\rangle\rangle}_{q} \enspace.
	\end{equation}

	Now using above relations one can prove that, in this case
	too, Tsallis 
	relative-entropy satisfies non-extensive triangle equality
	with modified conditions from the case of $q$-expectation
	values. We state it formally as follows. 
	Let $r$ be the prior estimate of the unknown distribution $l$
	which satisfies
	\begin{equation}
	\label{Equation:Constraint2_ForActualDistribution_Escort} 
	\frac{\int dx\, u_{m}(x) {l(x)}^{q} }{\int dx\, {l(x)}^{q}  }  =
	{\langle\langle{w}_{m}\rangle\rangle}_{q} \enspace, \:\:\: m =1, \ldots M 
	\enspace,
	\end{equation}
	where ${\langle\langle w_{m} \rangle\rangle}_{q}, \:\: m= 1,
	\ldots M$ are known normalized $q$-expected values of $l$. Let
	$p$ be the posterior which satisfies
	constraints~(\ref{Equation:EscortConstraint}). Then, similar to
	the calculations in
	\S~\ref{Section:NonextensiveTriangleEquality} one can prove
	that Tsallis relative entropy satisfies the nonextensive
	triangle equality
	(\ref{Equation:NonextensiveTriangleEquality}), provided
	\begin{equation}
	{\langle\langle u_{m}\rangle\rangle}_{q} = {\langle\langle
	w_{m}\rangle\rangle}_{q} \:\:\: m =1, \ldots M \enspace,
	\end{equation}
	but the minimum of $I_{q}(l\|p)$ is not guaranteed. Note that
	this condition is same as expectation value matching property
	in the classical case (see
	(\ref{Equation:ExpectedValueMatching}). 

	The detailed study of Tsallis relative-entropy minimization
	in this case of normalized $q$-expected values and the
	computation of corresponding minimum relative-entropy
	distribution (where one has to address the {\em
	self-referential} nature of the probabilities $p(x)$ in
	(\ref{Equation:Tsallis_minimumRelativeEntropyDistribution_For_Normalizedq-Expectations})) 
	based on Tsallis
	et. al~\cite{TsallisMendesPlastino:1998:TheRoleOfConstraints}, 
	Mart{\'\i}nez
	et.
	al~\cite{MartinezNicolasPenniniPlastino:2000:TsallisEntropyMaximizaionProcedureRevisited} 
	formalisms for Tsallis entropy maximization is under study.

\section{Conclusions}
\label{Section:Conclusion}

	Tsallis relative-entropy minimization has
	been studied and some significant differences with the
	classical case are presented. Generalized relative-entropy
	minimization has been shown to satisfy an {\em appropriate}
	generalized version of triangle equality for the classical 
	case. This is yet another remarkable and {\em consistant}
	generalization shown by Tsalls statistics.
	Considering the various fields to which Tsallis generalized
	statistics has been applied, studies of
	applications of Tsallis relative-entropy minimization for
	various inference problems are welcome.


\section*{Acknowledgments}
	The authors wish to thank anonymous refere for the 
	comments. The authors thank Prof. C. Tsallis for encouraging
	them to publish the present material upon going through the
	contents.  

\bibliographystyle{elsart-num}
\bibliography{papi}

\end{document}